\documentclass[10pt,a4paper]{article}
\usepackage{pifont}
\usepackage{amsmath}
\usepackage{amssymb}
\begin{document}
\title{Determining quark and lepton mass matrices by a geometrical
interpretation \\}

\author{Zhi-Qiang Guo\\
School of Physics, Peking University, Beijing 100871, China\\
\and Bo-Qiang Ma\footnote{Corresponding author; Electronic address:
mabq@phy.pku.edu.cn.} \\
School of Physics and MOE Key Laboratory of Heavy Ion Physics,
\\Peking University,
Beijing 100871, China\footnote{Mailing address.}\\
\date{}}

\maketitle

\begin{abstract}
   By designating one eigenvector of the mass matrix, one can reduce the free
parameters in the mass matrix effectively. Applying this method to
the quark mass matrix and to the lepton mass matrix, we find that
this method is consistent with available experimental data. This
approach may provide some hints for constructing theoretical models.
Especially, in the lepton sector, the Koide's mass relation is
connected to the element of the tribimaximal matrix through Foot's
geometrical interpretation. In the quark sector, we suggest another
mass formula and the same procedure also applies.
\end{abstract}

\vspace{3mm}

 I. INTRODUCTION

\vspace{3mm}

  The understanding of quark masses and mixings has posed a major challenge in
particle physics for a long time. Recently, the non-zero neutrino
masses and their mixings have been confirmed \cite{mohapatra}, which
implies that the mixing also exists in the lepton sector, just like
that in the quark sector. A key step to understand the masses and
mixings of quarks and leptons is to determine the mass matrices of
quarks and leptons. One popular method, suggested by Fritzsch
\cite{fritzsch}, is the texture zero structure. For example, the
four texture zero structure can survive current experimental tests
\cite{Fritzsch}. In the lepton sector, other matrices, for example,
based on the $\nu_\mu$-$\nu_\tau$ symmetry and the $A_4$ symmetry,
have been suggested \cite{mohapatra}. Especially, the
nearest-neighbor-interaction form~(NNI-form), can be implemented in
some grand unified theories \cite{Babu}, and is consistent with the
current experimental data from quarks and leptons \cite{Babu,Xing}.
Although the progress have been made in these directions, there is
still no a commonly granted standard theoretical model 
for these
problems. Therefore, other phenomenological approaches are necessary
and worthy to be explored, which may provide some hints for
constructing theoretical models. In this paper, we explore a way
that can realize Koide's mass formula \cite{koide} through Foot's
geometrical interpretation \cite{Foot}.

  The Yukawa sector of the standard model has too many free
parameters. In order to make definite predictions, we must make
efforts to reduce the redundant parameters effectively. As we have
emphasized, many papers have been devoted for this purpose. One
common character of these papers is to reduce the redundant
parameters by virtue of some symmetry \cite{mohapatra,Babu}. In this
paper, we explore another way, which is different from these
approaches. The main ideas are as follows. First, in the standard
model, we can choose the mass matrices to be Hermitian
\cite{branco}. Then by designating one eigenvector of a mass matrix,
we can reduce the redundant parameters effectively. In fact, as we
will illuminated below, if we make some assumptions and input the
values of the mass parameters, only four free parameters are left.
It is well known that the Cabibbo-Kobayashi-Maskawa~(CKM)
\cite{cabibbo} matrix has four parameters. Therefore, adjusting the
values of these left free parameters, we can fit the experimental
data in principle. We find that this way is consistent with
available experimental data. Because in this approach we have the
freedom to choose the eigenvectors, it gives us some advantage to
realize extra goals. For example, in the lepton sector, Koide's mass
formula is connected to the entry 33 of Maki-Nakagawa-Sakata~(MNS)
matrix \cite{maki} through Foot's geometrical interpretation, and it
is consistent with available experimental data. In the quark sector,
we suggested another mass formula, which can be connected to the
entry 33 of CKM matrix through Foot's geometrical interpretation.

  This paper is composed of five sections. In Sec.II, we introduce the method in detail. In Sec.III and IV, we apply this method to the
quark sector and to the lepton sector respectively. We make some
conclusions in Sec.V.

\vspace{5mm}
 II. THE METHOD
\vspace{3mm}

  In the standard model, the mass matrices are complex matrices in general, but
we can use the freedom of right-hand rotation to make them Hermitian
\cite{branco}. So without loss of generality, we start our
discussion from Hermite mass matrices.

   Supposed a general Hermite matrix
{\begin{eqnarray}
 \setcounter{equation}{1}\label{t1}
  \overline{M}=\left(\begin{array}{ccc}
  A & F\exp^{i\phi_F} & D\exp^{i\phi_D} \\
  F\exp^{-i\phi_F} & B & E\exp^{i\phi_E} \\
  D\exp^{-i\phi_D} &  E\exp^{-i\phi_E} & C
  \end{array}\right),
 \end{eqnarray}}
in which $A$, $B$, $C$, $\phi_D,~\phi_E,~\phi_F$ are real and $D$,
$E$, $F$ are nonnegative.

  The matrix $\overline{M}$ can be written in another way
{\begin{eqnarray}
 \setcounter{equation}{2}\label{t2}
  \overline{M}=P^{\dagger}MP=P^{\dagger}\left(\begin{array}{ccc}
  A & F & D \\
  F & B & E\exp^{i\alpha} \\
  D &  E\exp^{-i\alpha} & C
  \end{array}\right)P,
  \end{eqnarray}}
in which $P=\mathrm{diag}(1,\exp^{i\phi_F},\exp^{i\phi_D})$, and
$\alpha=\phi_E-\phi_D+\phi_F$. Given that $\overline{M}$ has one
eigenvector
 $\overrightarrow{x}^T=(x_1,x_2\exp^{i\beta},x_3\exp^{i\gamma})^T$
 belonging to its eigenvalue $\lambda$, we have the eigenequation
\begin{eqnarray}
\setcounter{equation}{3}\label{t3}
\overline{M}\overrightarrow{x}=\lambda\overrightarrow{x},
\end{eqnarray}
in which $x_1,~x_2$ and $x_3$ are nonnegative real numbers; $\beta$
and $\gamma$ are real numbers.

After some simplification, it reduces to
\begin{eqnarray}
 \setcounter{equation}{4}\label{t4}
 \left(\begin{array}{ccc}
  A-\lambda & F & D \\
  F & B-\lambda & E\exp^{i\alpha} \\
  D &  E\exp^{-i\alpha} & C-\lambda
  \end{array}\right)\left(\begin{array}{c}
  x_1 \\
  x_2\exp^{i(\beta+\phi_F)} \\
  x_3\exp^{i(\gamma+\phi_D)}
  \end{array}\right)=0.
  \end{eqnarray}
We see that Eq. (4) contains complex variables, and it will be
difficult to solve them. We notice that it will be simple in a
special case, in which we let $\beta=-\phi_F$ and $\gamma=-\phi_D$.
This implies that we designate a real eigenvector to the matrix $M$
in Eq. (\ref{t4}). By this choice, all of the equations have real
variables, and they can be solved with little labor. It is obviously
that this is only a conventional choice, which simplifies the
equation effectively. Of course, we should consider the possibility
that this choice may be not appropriate, hence the equations have no
solutions. However, in Sec.III and Sec.IV, we will give the
numerical results, which imply that our choice is compatible with
experimental data. In this paper, we will restrict our discussions
on this simple case. Of course, other cases, in which
$\beta+\phi_F\neq0$ and $\gamma+\phi_D\neq0$, are not excluded if
they are needed. Then we have
\begin{eqnarray}
 \setcounter{equation}{5}\label{t5}
\left(\begin{array}{ccc}
  A-\lambda & F & D \\
  F & B-\lambda & E\exp^{i\alpha} \\
  D &  E\exp^{-i\alpha} & C-\lambda
  \end{array}\right)\left(\begin{array}{c}
  x_1 \\
  x_2 \\
  x_3
  \end{array}\right)=0\Longleftrightarrow(M-\lambda I)\left(\begin{array}{c}
  x_1 \\
  x_2 \\
  x_3
  \end{array}\right)=0,
  \end{eqnarray}
in which $I$ is the identity matrix. In Eq. (\ref{t5}), as $A$, $B$,
$C$, $E$, $D$, $F$, $\lambda,~x_1,~x_2$ and $x_3$ are real numbers,
if E and $x_3$ are nonzero, $\alpha$ must equals to 0 or $\pi$.
Therefore, this matrix identity produces three equations. It is well
known that two of these equations are independent with each other.
We choose the independent equations to be
\begin{eqnarray}
\setcounter{equation}{6}\label{t6}
(A-\lambda)x_1+Fx_2+Dx_3&=&0,\\
\setcounter{equation}{7\label{t7}}
Fx_1+(B-\lambda)x_2+E\exp^{i\alpha}x_3&=&0,
\end{eqnarray}
in which $\alpha=0~\mathrm{or}~\pi$. In addition to Eq. (\ref{t6})
and Eq. (\ref{t7}), we have three eigenequations
\begin{eqnarray}
   \setcounter{equation}{8\label{t8}}
(A-\lambda_1)(B-\lambda_1)(C-\lambda_1)-E^2(A-\lambda_1)-D^2(B-\lambda_1)-F^2(C-\lambda_1)+2DEFN=0,\\
   \setcounter{equation}{9\label{t9}}
(A-\lambda_2)(B-\lambda_2)(C-\lambda_2)-E^2(A-\lambda_1)-D^2(B-\lambda_2)-F^2(C-\lambda_2)+2DEFN=0,\\
  \setcounter{equation}{10\label{t10}}
(A-\lambda_3)(B-\lambda_1)(C-\lambda_3)-E^2(A-\lambda_3)-D^2(B-\lambda_3)-F^2(C-\lambda_3)+2DEFN=0,
\end{eqnarray}
in which $N=\cos{\alpha}=\pm1$.

  So far, we have five equations, and we have six free
  parameters, among which only
one parameter is still free. We can let F to be the free parameter.
Once we fix the value of F, all other parameters are fixed. These
equations can be solved analytically, but the expressions are too
complicated. In order to simplify the expressions of the solutions,
we give some analysis in Appendix B. When we apply them to the
lepton sector in Sec.III and to the quark sector in Sec.IV, we will
give the analytical expressions explicitly in Appendix C and in
Appendix D. However, when we adjust the left free parameters to fit
the experimental data, the numerical approach is needed. In the
text, we display the numerical results. Also, it is possible that
these equations have no solutions, if the eigenvalue and the
eigenvector are not appropriate. However, in Sec.III and Sec.IV, we
will show that for the parameters we choose, the solutions always
exist, as we will display explicitly.

   The key point of our method is to choose the appropriate
eigenvalue and the appropriate eigenvector. In the following
application, we will choose the eigenvector and eigenvalue according
to physical ground.

   With the method we suggested above, if we fix the value of the left
free parameter, we can fix the matrix $M$. Now we turn to show how
we can use this method to determine the mixing matrix. We take the
CKM matrix for example. We let the up-quarks mass matrix to be
$\overline{M_u}$, and the down-quarks mass matrix to be
$\overline{M_d}$. Like $\overline{M}$, we can write $\overline{M_u}$
and $\overline{M_d}$ as
\begin{eqnarray}
   \setcounter{equation}{11\label{t11}}
\overline{M_u}=P_u^{\dagger}M_uP_u,~\overline{M_d}=P_d^{\dagger}M_dP_d.
\end{eqnarray}

  We designate
$\overrightarrow{y}^T=(y_1,y_2,y_3)^T$ as the eigenvector of $M_u$
belonging to its eigenvalue $\lambda_u$, and
$\overrightarrow{z}^T=(z_1,z_2,z_3)^T$ as the eigenvector of $M_d$
belonging to its eigenvalue $\lambda_d$. Then we have
\begin{eqnarray}
 \setcounter{equation}{12}\label{t12}
(M_u-\lambda_u I)\left(\begin{array}{c}
  y_1 \\
  y_2 \\
  y_3
  \end{array}\right)=0,~~
(M_d-\lambda_d I)\left(\begin{array}{c}
  z_1 \\
  z_2 \\
  z_3
  \end{array}\right)=0.
\end{eqnarray}
  According to our analysis above, all the elements of $M_u$ and $M_d$
can be determined except two of them. Suppose that we input the
values of these two free parameters, then we can determine $M_u$ and
$M_d$. $M_u$ and $M_d$ can be diagonalized by orthogonal
transformation
\begin{eqnarray}
   \setcounter{equation}{13\label{t13}}
V_u^TM_uV_u=\mathrm{diag}(m_u,m_c,m_t),~
V_d^TM_dV_d=\mathrm{diag}(m_d,m_s,m_b).
\end{eqnarray}
By Eq. (\ref{t11}), Eq. (\ref{t13}) can be rewritten as
\begin{eqnarray}
   \setcounter{equation}{14\label{t14}}
V_u^TP_u\overline{M_u}P_u^{\dagger}V_u=\mathrm{diag}(m_u,m_c,m_t),~
V_d^TP_d\overline{M_d}P_d^{\dagger}V_d=\mathrm{diag}(m_d,m_s,m_b).
\end{eqnarray}
The CKM matrix can be defined as
\begin{eqnarray}
   \setcounter{equation}{15\label{t15}}
V_{CKM}=V_u^TP_uP_d^{\dagger}V_d.
\end{eqnarray}

By our analysis above,
$P=P_uP_d^{\dagger}=\mathrm{diag}(1,\exp^{i\xi},\exp^{i\eta}).$
Therefore, we have four free parameters, i.e., $F_u$ and $F_d$
respectively in $M_u$ and $M_d$, and $\xi$ and $\eta$ in $P$. It is
well known that the CKM matrix have four free parameters. Hence in
principle it is possible that we can adjust the values of our free
parameters to make them consistent with the experimental data. The
similar procedure also applies to the lepton sector.

\vspace{5mm}
 III. THE APPLICATION TO THE LEPTON SECTOR
\vspace{3mm}

   In recent years, neutrino physics has made great progress. The mixing of neutrinos
has been confirmed and the structure of neutrino mixing matrix has
been determined to a reasonable precision \cite{mohapatra}. It is
well known that the neutrino mixing matrix can be expressed with the
tribimaximal matrix approximately \cite{Harrison}. It is
\begin{eqnarray}
   \setcounter{equation}{16\label{t16}}
  V_{MNS}=V_{l L}^{\dagger}V_{\nu L}=\left(\begin{array}{ccc}
  \sqrt{6}/3 & \sqrt{3}/3 & 0 \\
  -\sqrt{6}/6& \sqrt{3}/3 & \sqrt{2}/2 \\
  \sqrt{6}/6 & -\sqrt{3}/3 & \sqrt{2}/2
  \end{array}\right)
  \end{eqnarray}

  The important step of our method is to choose the eigenvectors
appropriately. In order to choose eigenvectors in the lepton
sector, we notice some investigations below.

  In the lepton sector, Koide \cite{koide} ever suggested an accurate formula
\begin{eqnarray}
   \setcounter{equation}{17\label{t17}}
  \frac{1}{\sqrt{2}}=\frac{\sqrt{m_e}+\sqrt{m_\mu}+\sqrt{m_\tau}}{\sqrt{3}\sqrt{m_e+m_\mu+m_\tau}}.
  \end{eqnarray}

  Foot \cite{Foot} gave a geometrical interpretation to it,
\begin{eqnarray}
   \setcounter{equation}{18\label{t18}}
  \cos{\theta}=\frac{(\sqrt{m_e},\sqrt{m_\mu},\sqrt{m_\tau})(1,1,1)}{\mid(\sqrt{m_e},\sqrt{m_\mu},\sqrt{m_\tau})\mid\mid
  (1,1,1)\mid},
  \end{eqnarray}
where$(\sqrt{m_e},\sqrt{m_\mu},\sqrt{m_\tau})$ and $(1,1,1)$ are
interpreted as two vectors, and $\theta$ is the angle between them.
If we choose $\theta=\frac{\pi}{4}$, we get the Koide's mass
formula. We have seen that in the tribimaximal matrix, the entry 33
is approximate to be $\frac{1}{\sqrt{2}}$. Therefore there is a
natural connection between them. The Koide's mass formula and Foot's
geometrical interpretation provide hints for choosing the
eigenvectors.

  We choose $(\sqrt{m_e},\sqrt{m_\mu},\sqrt{m_\tau})$ and $(1,1,1)$ as
the eigenvectors we want. It is proper that we choose $(1,1,1)$ as
the eigenvector of neutrino mass matrix belonging to $m_3$~(the mass
of a neutrino) and $(\sqrt{m_e},\sqrt{m_\mu},\sqrt{m_\tau})$ as the
eigenvector of the charged lepton matrix belonging to $m_\tau$. This
implies that we designate the vectors as follows
\begin{eqnarray}
   \setcounter{equation}{19\label{t19}}
  V_{MNS}=V_{l L}^{\dagger}V_{\nu L}=\left(\begin{array}{ccc}
   \star &  \star &  \star\\
   \star&  \star & \star \\
  \frac{\sqrt{m_e}}{\sqrt{m_e+m_{\mu}+m_{\tau}}}&\frac{\sqrt{m_{\mu}}}{\sqrt{m_e+m_{\mu}+m_{\tau}}} &\frac{\sqrt{m_{\tau}}}{\sqrt{m_e+m_{\mu}+m_{\tau}}}
  \end{array}\right)P\left(\begin{array}{ccc}
   \star &  \star & \frac{1}{\sqrt{3}}\\
   \star&  \star & \frac{1}{\sqrt{3}} \\
  \star & \star & \frac{1}{\sqrt{3}}
  \end{array}\right).
  \end{eqnarray}
where $P=\mathrm{diag}(1,\exp^{i\xi},\exp^{i\eta}).$

  As we described above, we need the values of the mass parameters. For the
charged leptons, the masses are known accurately. However, for the
neutrinos, only the mass squared differences are measured.  The
results of global analysis read \cite{mohapatra,Strumia}
\begin{eqnarray}
\setcounter{equation}{20\label{t20}} \triangle
m_{12}^2=m_2^2-m_1^2=(7.9\pm0.4)\times 10^{-5}~\mathrm{eV^2},~(1\sigma)\\
\setcounter{equation}{21\label{t21}} \mid\triangle m_{32}^2\mid=\mid
m_3^2-m_2^2\mid=(2.4\pm0.3)\times 10^{-3}~\mathrm{eV^2}.~(1\sigma)
\end{eqnarray}

If we assume the normal mass hierarchy, i.e.,
\begin{eqnarray}
\setcounter{equation}{22\label{t22}} m_3>m_2>m_1,
\end{eqnarray}
then we have
\begin{eqnarray}
\setcounter{equation}{23\label{t23}} \frac{m_{21}^2}{m_{31}^2}\cong
\frac{m_{21}^2}{m_{23}^2}=0.033\pm0.004,\\
\setcounter{equation}{24\label{t24}}
\frac{m_2}{m_3}\geqslant\sqrt{\frac{m_{21}^2}{m_{31}^2}}=0.18\pm0.01.
\end{eqnarray}

  It is obvious that the mixing matrix only depends on the
mass ratio \cite{jarlskog}. So we only need the ratios of neutrinos
masses. It is convenient to normalize the neutrinos masses as
follows
\begin{eqnarray}
\setcounter{equation}{25\label{t25}}
\lambda_1=m_1=0.1m_3,~\lambda_2=m_2=0.2m_3,~\lambda_3=m_3.
\end{eqnarray}
Note that these values of mass parameters do not stand for the
absolute mass, but just stand for the mass ratio. These ratios are
consistent with the experimental data we have displayed above.

Let $x=1$, $y=1$ and $F=0.279853~m_3$ in the neutrino mass matrix,
by Eqs. (\ref{eq8}), (\ref{eq9}), (\ref{eq10}), (\ref{eq13}) and
(\ref{eq15}) in Appendix C and Appendix D. we can get the neutrino
mass matrix
\begin{eqnarray}
\setcounter{equation}{26\label{t26}}
  M_\nu=m_3\left(\begin{array}{ccc}
  0.463783 &  0.279853 & 0.256364\\
  0.279853 &  0.406364 & 0.313783  \\
 0.256364& 0.313783 & 0.429853
  \end{array}\right).
\end{eqnarray}

For the charged leptons, let
\begin{eqnarray}
\setcounter{equation}{27\label{t27}}
\lambda_1=m_e=0.511~\mathrm{MeV},~\lambda_2=m_\mu=105.658~\mathrm{MeV},~\lambda_3=m_\tau=1776.97~\mathrm{MeV},
\end{eqnarray}
and
\begin{eqnarray}
\setcounter{equation}{28\label{t28}}
x=\frac{\sqrt{m_e}}{\sqrt{m_\tau}},~y=\frac{\sqrt{m_\mu}}{\sqrt{m_\tau}},~F=55.6898~\mathrm{MeV},
\end{eqnarray}
similarly we can get the charged leptons mass matrix
\begin{eqnarray}
\setcounter{equation}{29\label{t29}}
  M_l=\left(\begin{array}{ccc}
  68.2912 &  55.6898 & 15.3959\\
  55.6898 &  135.509 & 399.315  \\
 15.3959& 399.315 &1679.34
  \end{array}\right)~\mathrm{MeV}.
  \end{eqnarray}

Note that in the course of getting the mass matrices $M_l$ and
$M_\nu$ above, we have chosen the eigenvectors of $M_l$ and $M_\nu$
to be real, hence the mass matrices are real matrices according to
our analysis in Sec.II. Here we emphasize that this just is a
conventional choice, which simplifies the equations effectively.

The matrices that diagonalizes $M_l$ and $M_\nu$ are given as
\begin{eqnarray}
\setcounter{equation}{30\label{t30}}
  V_l=\left(\begin{array}{ccc}
  0.599741& -0.800024 & 0.0164729\\
  -0.779697& -0.579625 &0.23687 \\
 0.179954&0.154905& 0.971402
  \end{array}\right),~
  V_\nu=\left(\begin{array}{ccc}
   0.169807& -0.798644 & 0.57735 \\
  -0.776549& 0.252265 & 0.57735 \\
 0.606743 & 0.546379 & 0.57735
  \end{array}\right).
  \end{eqnarray}
Obviously the third column of $V_l$ is
$(\frac{\sqrt{m_e}}{\sqrt{m_e+m_{\mu}+m_{\tau}}},\frac{\sqrt{m_\mu}}{\sqrt{m_e+m_{\mu}+m_{\tau}}},\frac{\sqrt{m_\tau}}{\sqrt{m_e+m_{\mu}+m_{\tau}}})^T$,
and the third column of $V_\nu$ is
$(\frac{1}{\sqrt{3}},\frac{1}{\sqrt{3}},\frac{1}{\sqrt{3}})^T$.

  The MNS matrix is given as
\begin{eqnarray}
\setcounter{equation}{31\label{t31}}
  V_{MNS}=V_l^TP_lP_{\nu}^{\dagger}V_{\nu}P=V_l^T\left(\begin{array}{ccc}
  1&0& 0\\
  0&\exp^{i\xi} &0\\
 0&0&\exp^{i\eta}
  \end{array}\right)V_{\nu}\left(\begin{array}{ccc}
  \exp^{i\alpha_1}&0& 0\\
  0&\exp^{i\alpha_2} &0\\
  0&0&1
  \end{array}\right).
  \end{eqnarray}

The phases $\alpha_1$ and $\alpha_2$, known as the Majorana phases,
have physical consequences only if neutrinos are Majorana particles.
Because there is no clear evidence whether there is CP-violation in
the lepton sector, we can not fix the values of $\xi$ and $\eta$.
Once we can measure the MNS matrix accurately, we can adjust the
free parameters $F,~\xi$ and $\eta$ to fit the experimental data. If
we let $\xi=0$ and $\eta=0$, we obtain
\begin{eqnarray}
\setcounter{equation}{32\label{t32}}
  V_{MNS}=\left(\begin{array}{ccc}
  0.816499&-0.577347& 0\\
  0.408246&0.577352&-0.707107\\
 0.408247&0.577352&0.707107
  \end{array}\right)\left(\begin{array}{ccc}
  \exp^{i\alpha_1}&0& 0\\
  0&\exp^{i\alpha_2} &0\\
  0&0&1
  \end{array}\right).
  \end{eqnarray}
The magnitude of the elements are given as
\begin{eqnarray}
\setcounter{equation}{33\label{t33}}
  \mid V_{MNS}\mid=\left(\begin{array}{ccc}
  0.816499&0.577347& 0\\
  0.408246&0.577352&0.707107\\
 0.408247&0.577352&0.707107
  \end{array}\right)
 ,
  \end{eqnarray}
which is very close to the tribimaximal matrix.

\vspace{5mm}
 IV. THE APPLICATION TO THE QUARK SECTOR
\vspace{3mm}

   The quark mixing matrix has been determined in high
precision. The CKM matrix elements can be most precisely determined
by a global fit that uses all available measurements and imposes the
standard model constraints. The allowed ranges of the magnitudes of
all CKM elements are \cite{yao}
\begin{eqnarray}
\setcounter{equation}{34\label{t34}}
 V_{CKM}=V_{uL}^{\dagger}V_{dL}=\left(\begin{array}{ccc}
  0.97360~\mathrm{to}~0.97407 & 0.2262~\mathrm{to}~0.2282 &0.00387~\mathrm{to}~0.00405\\
  0.2261~\mathrm{to}~0.2281&0.97272~\mathrm{to}~0.97320&0.04141~\mathrm{to}~0.04231\\
  0.0075~\mathrm{to}~0.00846& 0.04083~\mathrm{to}~0.04173
  &0.999096~\mathrm{to}~0.999134
  \end{array}\right).
  \end{eqnarray}

   Just like that in the lepton sector, we also need two vectors. We find that a
numerical relation is well satisfied. It reads
\begin{eqnarray}
\setcounter{equation}{35\label{t35}}
  \cos{\theta}=\frac{(\sqrt{m_u},\sqrt{m_c},\sqrt{m_t})(m_d,m_s,m_b)}{\mid(\sqrt{m_u},\sqrt{m_c},\sqrt{m_t})\mid\mid(m_d,m_s,m_b)
  \mid}=0.999549.
  \end{eqnarray}
The reasons that we choose this formula are displayed in Appendix A.

Hence we might speculate that $\theta=0$, while in the Koide's mass
relation $\theta=\frac{\pi}{4}$. Because the elements of $V_{CKM}$
are given at the scale $\mu=M_Z$, we use the quark mass given at the
scale $\mu=M_Z$.  This well satisfied numerical relation suggests us
to designate the vectors as follows
\begin{eqnarray}
\setcounter{equation}{36\label{t36}}
 V_{CKM}=V_{uL}^{\dagger}V_{dL}=\left(\begin{array}{ccc}
   \star &  \star &  \star\\
   \star&  \star & \star \\
  \frac{\sqrt{m_u}}{\sqrt{m_u+m_c+m_t}}& \frac{\sqrt{m_c}}{\sqrt{m_u+m_c+m_t}}&\frac{\sqrt{m_t}}{\sqrt{m_u+m_c+m_t}}
  \end{array}\right)P\left(\begin{array}{ccc}
   \star &  \star & \frac{m_d}{\sqrt{m_d^2+m_s^2+m_b^2}}\\
   \star&  \star & \frac{m_s}{\sqrt{m_d^2+m_s^2+m_b^2}} \\
  \star & \star & \frac{m_b}{\sqrt{m_d^2+m_s^2+m_b^2}}
  \end{array}\right),
  \end{eqnarray}
where $P=\mathrm{diag}(1,\exp^{i\xi},\exp^{i\eta}).$

  It implies that the vector $(\frac{\sqrt{m_u}}{\sqrt{m_u+m_c+m_t}},\frac{\sqrt{m_c}}{\sqrt{m_u+m_c+m_t}},\frac{\sqrt{m_t}}{\sqrt{m_u+m_c+m_t}})^T$ is the eigenvector of $M_u$ belonging to its eigenvalue
  $m_t$, and the vector $(\frac{m_d}{\sqrt{m_d^2+m_s^2+m_b^2}},\frac{m_s}{\sqrt{m_d^2+m_s^2+m_b^2}},\frac{m_b}{\sqrt{m_d^2+m_s^2+m_b^2}})^T$ is the eigenvector of $M_d$ belonging to its eigenvalue
  $m_b$.
  As we have emphasized before, if we fix the value of the free parameter,
 other elements of the mass matrix are determined. The equations can
be solved analytically. We display the result in Appendix C.

  For the up-quark sector, let
\begin{eqnarray}
\setcounter{equation}{37\label{t37}}
\lambda_1=m_u(M_Z)=2.33~\mathrm{MeV},~\lambda_2=m_c(M_Z)=677~\mathrm{MeV},~\lambda_3=m_t(M_Z)=181000~\mathrm{MeV},
\end{eqnarray}
 and
\begin{eqnarray}
\setcounter{equation}{38\label{t38}}
 x=\frac{\sqrt{m_u}}{\sqrt{m_t}},~y=\frac{\sqrt{m_c}}{\sqrt{m_t}},~F=133.18~\mathrm{MeV},
\end{eqnarray}
we can get the up-quark mass matrix by Eqs. (\ref{eq8}),
(\ref{eq9}), (\ref{eq10}), (\ref{eq13}) and (\ref{eq14}) in Appendix
C.
\begin{eqnarray}
\setcounter{equation}{39\label{t39}}
  M_u=\left(\begin{array}{ccc}
  17.9519& 133.18 & 641.198\\
  133.18&  1335.65 &10987.5  \\
 641.198& 10987.5 & 180326
  \end{array}\right)~\mathrm{MeV}.
\end{eqnarray}

  For the down-quark sector, let
\begin{eqnarray}
\setcounter{equation}{40\label{t40}}
\lambda_1=m_d(M_Z)=4.69~\mathrm{MeV},~\lambda_2=m_s(M_Z)=93.4~\mathrm{MeV},~\lambda_3=m_b(M_Z)=3000~\mathrm{MeV},
\end{eqnarray} and
\begin{eqnarray}
\setcounter{equation}{41\label{t41}}
x=\frac{m_d}{m_b},~y=\frac{m_s}{m_b},~F=7.99~\mathrm{MeV},
\end{eqnarray}
 similarly we can get the down-quark mass matrix
\begin{eqnarray}
\setcounter{equation}{42\label{t42}}
M_d=\left(\begin{array}{ccc}
   5.39717& 7.99 &4.43281 \\
  7.99& 95.5146 & 90.4138 \\
 4.43281& 90.4138 &2997.18
  \end{array}\right)~\mathrm{MeV}.
\end{eqnarray}

Again, in the course of getting the mass matrices $M_u$ and $M_d$,
for simplicity, we choose the eigenvectors to be real, hence the
mass matrices are real matrices by the analysis in Sec.II.

The matrices that diagonalize $M_u$ and $M_d$ are given as
\begin{eqnarray}
\setcounter{equation}{43\label{t43}}
  V_u=\left(\begin{array}{ccc}
  0.990087&-0.140407 & 0.00358117\\
  -0.140363& -0.988217 &0.0610438\\
 0.00503202&0.0609414 & 0.998129
  \end{array}\right),~
  V_d=\left(\begin{array}{ccc}
   0.996046& -0.0888212&0.00156257 \\
  -0.0888268& -0.995561&0.0311182 \\
 0.00120832&0.031134 &0.999514
  \end{array}\right).
  \end{eqnarray}
Obviously the third column of $V_u$ is
$(\frac{\sqrt{m_u}}{\sqrt{m_u+m_c+m_t}},\frac{\sqrt{m_c}}{\sqrt{m_u+m_c+m_t}},\frac{\sqrt{m_t}}{\sqrt{m_u+m_c+m_t}})^T$,
and the third column of $V_d$ is
$(\frac{m_d}{\sqrt{m_d^2+m_s^2+m_b^2}},\frac{m_s}{\sqrt{m_d^2+m_s^2+m_b^2}},\frac{m_b}{\sqrt{m_d^2+m_s^2+m_b^2}})^T
$.

The CKM matrix is given as
\begin{eqnarray}
\setcounter{equation}{44\label{t44}}
  V_{CKM}=V_u^TP_uP_d^{\dagger}V_d=V_u^T\left(\begin{array}{ccc}
  1&0& 0\\
  0&\exp^{i\xi} &0\\
 0&0&\exp^{i\eta}
  \end{array}\right)V_d.
  \end{eqnarray}

Given $\xi=0.76\pi+\frac{\pi}{4.515}$ and $\eta=0.76\pi$, the CKM
matrix equals to
\begin{eqnarray}
\setcounter{equation}{45\label{t45}}
V_{CKM}=\left(\begin{array}{ccc}
   0.973722+0.000729014i&-0.227558+0.00823131i &0.00224113+0.00318905i \\
  -0.227538+0.00515369i&-0.971078+0.0584958i &-0.0139226+0.0399092i \\
  0.00810099+0.000510366i&0.0376987+0.0177396i&-0.729142+0.683045i
  \end{array}\right).
  \end{eqnarray}

The magnitude of the elements are
\begin{eqnarray}
\setcounter{equation}{46\label{t46}}
\mid V_{CKM}
\mid=\left(\begin{array}{ccc}
  0.973722&0.227707&0.00389778 \\
 0.227596&0.972838 &0.042268 \\
  0.00811705&0.041664&0.999099
  \end{array}\right).
  \end{eqnarray}

The quantities of rephasing invariance are calculated as
\begin{eqnarray}
\setcounter{equation}{47\label{t47}}
\alpha=\phi_2=\mathrm{arg}[-\frac{V_{td}V_{tb}^*}{V_{ud}V_{ub}^*}]=101.59^\circ,\beta=\phi_1=\mathrm{arg}[-\frac{V_{cd}V_{cb}^*}{V_{td}V_{tb}^*}]=22.74^\circ,\gamma=\phi_3=\mathrm{arg}[-\frac{V_{ud}V_{ub}^*}{V_{cd}V_{cb}^*}]=55.67^\circ,
\end{eqnarray}
and the invariant measure of CP violation is calculated as
\begin{eqnarray}
\setcounter{equation}{48\label{t48}}
J=-\mathrm{Im}(V_{ud}V_{cb}V_{ub}^*V_{cd}^*)=3.01513\times10^{-5},
\end{eqnarray}
 while the the experimental data \cite{yao} are given as
\begin{eqnarray}
\setcounter{equation}{49\label{t49}}
\alpha=(99^{+13}_{-8})^\circ,~~\beta=(21.70^{+1.29}_{-1.24})^\circ,~~\gamma=(63^{+15}_{-12})^\circ,~~J=(3.08^{+0.16}_{-0.18})\times10^{-5}.
\end{eqnarray}
They are consistent with each other.

\vspace{5mm}

V. CONCLUSION

\vspace{3mm}

  We have illustrated the method in Sec.II, and apply it to the lepton
sector in Sec.III and to the quark sector in Sec.IV respectively.
The character of this method is to designate the eigenvector and the
eigenvalue for the mass matrix appropriately. In the lepton sector,
we use the Koide's formula, and in the quark sector, we use a
similar formula that is well satisfied. Now we give some comments
about the mass formula we used.

(1) In the lepton sector, the Koide's mass formula,
\begin{eqnarray}
\setcounter{equation}{50\label{t50}}
  \frac{1}{\sqrt{2}}=\frac{\sqrt{m_e}+\sqrt{m_\mu}+\sqrt{m_\tau}}{\sqrt{3}\sqrt{m_e+m_\mu+m_\tau}},
\end{eqnarray}
is satisfied in high precision. It is energy scale insensitive
\cite{Li}, and its other characters were also discussed
\cite{Gerard}. Several explanations that can realize this formula
have been given \cite{Koide}. Among these explanations, Foot's
geometrical interpretation seems fascinating phenomenologically.
However, there is still no a theoretical model that can realize it.
It seems that our method can implement this interpretation. If there
is no CP-violation and the MNS matrix is tribimaximal, the righthand
of the mass formula is connected to the element of the MNS matrix by
Foot's geometrical interpretation. This is an approach that can lead
to the Koide's mass formula.

(2) In the quark sector, the mass formula like Koide's is
unsuccessful \cite{Espoeito}. However, because the CKM matrix is
very different from the MNS matrix, such a mass formula is useless
for us. Alternately, we find another well satisfied mass formula,
\begin{eqnarray}
\setcounter{equation}{51\label{t51}}
  \cos{\theta}=\frac{(\sqrt{m_u},\sqrt{m_c},\sqrt{m_t})(m_d,m_s,m_b)}{\mid(\sqrt{m_u},\sqrt{m_c},\sqrt{m_t})\mid\mid(m_d,m_s,m_b)
  \mid}=0.999549.
\end{eqnarray}
  This mass formula is connected to the element of the CKM
matrix.

(3) There are two reasons that we choose the mass formula as the
vectors. First, the vectors are expressed by the mass parameters, so
we do not need to introduce extra parameters. Obviously that this is
an economic choice. Second, there exists such mass formula. They are
excellent and are well satisfied in high precision, like the Koide's
mass formula, but we can not realize them in a concise and
convincing way. Our method provides an approach that can realize
them, nevertheless in the special situation if there is no
CP-violation. Of course other choices of the vectors are also
permitted.

  Finally we give some comments about the texture zero structure and our
method. The differences between our method and the texture zero
structure are: in our present approach, at first we choose one
eigenvector of the mass matrix, and then we can determine other
elements of the mass matrix; in the texture zero structure, some
elements of the mass matrix are supposed to zero, which equals to
designate the eigenvectors, but we do not know the eigenvector at
first. Therefore, in our approach, we have the freedom to choose the
eigenvectors to satisfy other request. For example, the Koide's mass
formula can be realized in our approach through Foot's geometrical
interpretation.

   As the texture zero structure, our approach is also consistent
with current experimental data, and this approach has the merit that
it can realize some well satisfied mass formula, for example, the
Koide's mass formula and the mass formula suggested by us. However,
just as many texture zero structures, our approach is purely
phenomenological, and there is still no a theoretical model to
realize it. Therefore it is worthy to investigate whether our
approach can be realized in some theoretical models. If this is
true, it will provide a new theoretical and phenomenological
approach to deal with the mass and mixing problems. Besides, we
point out that in our paper we just restrict our discussions in a
simple case, in which we choose the eigenvectors to be real, hence
we get the real mass matrices. This is just a conventional choice
that simplifies the equations. Other cases, in which the
eigenvectors are complex, are permitted, and they should be
considered if they are needed by some underlying theories.

{\bf Acknowledgement} This work is partially supported by National
Natural Science Foundation of China (Nos.~10421503, 10575003,
10528510), by the Key Grant Project of Chinese Ministry of Education
(No.~305001), and by the Research Fund for the Doctoral Program of
Higher Education (China).

\begin{appendix}
\vspace{9mm}
 APPENDIX A
\vspace{3mm}

Table: Quark masses at the Z mass scale in the standard model
\cite{Fusaoka} \vspace{1mm}

\begin{tabular}{ll}\hline
 $m_u(M_Z)=2.33_{-0.45}^{+0.42}~\mathrm{MeV}$&
$m_d(M_Z)=4.69_{-0.66}^{+0.60}~\mathrm{MeV}$\\
$m_c(M_Z)=677_{-61}^{+56}~\mathrm{MeV}$&
$m_s(M_Z)=93.4_{-13.0}^{+11.8}~\mathrm{MeV}$\\
 $m_t(M_Z)=181_{-13}^{+13}~\mathrm{GeV}$&
$m_b(M_Z)=3.00_{-0.11}^{+0.11}~\mathrm{GeV}$\\
\hline
\end{tabular}
\vspace{1mm}

With the values of the quark masses above,
$$
m_u(M_Z)=2.33~\mathrm{MeV},~m_c(M_Z)=677~\mathrm{MeV},
~m_t(M_Z)=181~\mathrm{GeV},$$
$$m_d(M_Z)=4.69~\mathrm{MeV},~m_s(M_Z)=93.4~\mathrm{MeV},~m_b(M_Z)=3.00~\mathrm{GeV},$$
\[
  \cos{\theta}=\frac{(\sqrt{m_u},\sqrt{m_c},\sqrt{m_t})(m_d,m_s,m_b)}{\mid(\sqrt{m_u},\sqrt{m_c},\sqrt{m_t})\mid\mid(m_d,m_s,m_b)
  \mid}=0.999549,
  \]
$\cos{\theta}=0.999549$ is very close to 1.

   Note that we do not choose the mass formula below,
   \[
  \cos{\theta}=\frac{(m_u,m_c,m_t)(\sqrt{m_d},\sqrt{m_s},\sqrt{m_b})}{\mid(m_u,m_c,m_t)\mid\mid(\sqrt{m_d},\sqrt{m_s},\sqrt{m_b})
  \mid}=0.984685,
  \]
\[
  \cos{\theta}=\frac{(m_u,m_c,m_t)(m_d,m_s,m_b)}{\mid(m_u,m_c,m_t)\mid\mid(m_d,m_s,m_b)
  \mid}=0.999624,
  \]
\[
  \cos{\theta}=\frac{(\sqrt{m_u},\sqrt{m_c},\sqrt{m_t})(\sqrt{m_d},\sqrt{m_s},\sqrt{m_b})}{\mid(\sqrt{m_u},\sqrt{m_c},\sqrt{m_t})\mid\mid(\sqrt{m_d},\sqrt{m_s},\sqrt{m_b})
  \mid}=0.992936.
  \]
   As we have shown in Sec.IV, the mass formula is
connected to the element of the CKM matrix by,
\[
V_{33}=
\frac{m_u\sqrt{m_d}+\exp^{i\xi}m_c\sqrt{m_s}+\exp^{i\eta}m_t\sqrt{m_b}}{\sqrt{m_u^2+m_c^2+m_t^2}\sqrt{m_d+m_s+m_b}},
\]
\[
  V_{33}=\frac{m_u m_d+\exp^{i\xi}m_c m_s+\exp^{i\eta}m_t
  m_b}{\sqrt{m_u^2+m_c^2+m_t^2}\sqrt{m_d^2+m_s^2+m_b^2}},
  \]
\[
V_{33}=
\frac{\sqrt{m_u}\sqrt{m_d}+\exp^{i\xi}\sqrt{m_c}\sqrt{m_s}+\exp^{i\eta}\sqrt{m_t}\sqrt{m_b}}{\sqrt{m_u+m_c+m_t}\sqrt{m_d+m_s+m_b}},
\]
and
\[
0.983385\leqslant\mid
\frac{m_u\sqrt{m_d}+\exp^{i\xi}m_c\sqrt{m_s}+\exp^{i\eta}m_t\sqrt{m_b}}{\sqrt{m_u^2+m_c^2+m_t^2}\sqrt{m_d+m_s+m_b}}\mid\leqslant0.984685,
\]
\[
  0.999391\leqslant\mid\frac{m_u m_d+\exp^{i\xi}m_c m_s+\exp^{i\eta}m_t
  m_b}{\sqrt{m_u^2+m_c^2+m_t^2}\sqrt{m_d^2+m_s^2+m_b^2}}\mid\leqslant0.999624,
  \]
\[
0.971465\leqslant\mid
\frac{\sqrt{m_u}\sqrt{m_d}+\exp^{i\xi}\sqrt{m_c}\sqrt{m_s}+\exp^{i\eta}\sqrt{m_t}\sqrt{m_b}}{\sqrt{m_u+m_c+m_t}\sqrt{m_d+m_s+m_b}}\mid\leqslant0.992936,
\]
but the experimental value of $\mid V_{33}\mid$ is
\[
0.999096<\mid V_{33}\mid<0.999134.
\]
They are not consistent. Therefore, in order to make them consistent
with the experimental data, we must change the mass parameters that
we have used. But the mass formula we have chosen does not have this
problem. Our choice is just a convenient one. Of course other
choices are permitted if they are consistent with the experimental
data.

\vspace{5mm}
 APPENDIX B
\vspace{3mm}

  As we have emphasized in the text, the analytical expressions of the
solutions will be complicated. In this appendix, we give some
analysis that will simplify the expressions effectively.

  In order to express the other elements of the mass
matrix with  the quarks masses and the free parameter $F$, we have
to solve the equations displayed below
\begin{eqnarray}
\setcounter{equation}{1\label{eq1}}
(A-\lambda)x_1+Fx_2+Dx_3=0,\\
\setcounter{equation}{2\label{eq2}}
 Fx_1+(B-\lambda)x_2+E\exp^{i\alpha}x_3=0,\\
 \setcounter{equation}{3\label{eq3}}
(A-\lambda_1)(B-\lambda_1)(C-\lambda_1)-E^2(A-\lambda_1)-D^2(B-\lambda_1)-F^2(C-\lambda_1)+2DEFN=0,\\
   \setcounter{equation}{4\label{eq4}}
(A-\lambda_2)(B-\lambda_2)(C-\lambda_2)-E^2(A-\lambda_2)-D^2(B-\lambda_2)-F^2(C-\lambda_2)+2DEFN=0,\\
  \setcounter{equation}{5\label{eq5}}
(A-\lambda_3)(B-\lambda_3)(C-\lambda_3)-E^2(A-\lambda_3)-D^2(B-\lambda_3)-F^2(C-\lambda_3)+2DEFN=0.
\end{eqnarray}

It is well known that
\begin{eqnarray}
\setcounter{equation}{6\label{eq6}}
\mathrm{Trace}(M)=\lambda_1+\lambda_2+\lambda_3\Longrightarrow
A+B+C=s,
\end{eqnarray}
in which $s=\lambda_1+\lambda_2+\lambda_3.$

If $\lambda_1\neq \lambda_2$, by Eqs. (\ref{eq3}), (\ref{eq4}) and
(\ref{eq6}), we obtain
\begin{eqnarray}
\setcounter{equation}{7\label{eq7}} AB+BC+AC=r+(D^2+E^2+F^2),
\end{eqnarray}
in which
$r=\lambda_1\lambda_3+\lambda_1\lambda_2+\lambda_2\lambda_3.$

By Eqs. (\ref{eq1}) and (\ref{eq2}), we can express $D$ and $E$ in
terms of $A$, $B$ and $F$
\begin{eqnarray}
\setcounter{equation}{8\label{eq8}}
 D=-(A-\lambda)x-Fy,\\
 \setcounter{equation}{9\label{eq9}}
 E=[-Fx-(B-\lambda)y]\exp^{-i\alpha},
\end{eqnarray}
in which we have let $x=\frac{x_1}{x_3}$ and $y=\frac{x_2}{x_3}$,
supposing that $x_3\neq 0$.

We obtain five new equations
\begin{eqnarray}
\setcounter{equation}{5\label{eq5}}
(A-\lambda_3)(B-\lambda_3)(C-\lambda_3)-E^2(A-\lambda_3)-D^2(B-\lambda_3)-F^2(C-\lambda_3)+2DEFN=0,\\
\setcounter{equation}{6\label{eq6}}
A+B+C=s,\\
\setcounter{equation}{7\label{eq7}}
AB+BC+AC=r+(D^2+E^2+F^2),\\
\setcounter{equation}{8\label{eq8}}
 D=-(A-\lambda)x-Fy,\\
 \setcounter{equation}{9\label{eq9}}
 E=[-Fx-(B-\lambda)y]\exp^{-i\alpha}.
\end{eqnarray}
Let $N=\cos{\alpha}=\pm 1$, by these equations, we can express $A$,
$B$, $C$, $D$ and $E$ in terms of $F$.

In the following we will deduce another formula, which will simplify
the expressions effectively. By Eq. (\ref{eq6}), we obtain
\begin{eqnarray}
\setcounter{equation}{10\label{eq10}} C=s-A-B.
\end{eqnarray}

We have argued that $N=\cos{\alpha}=\pm 1$ in Sec.II. So by Eq.
(\ref{eq9}), we have
\begin{eqnarray}
\setcounter{equation}{11\label{eq11}} E=\pm [-Fx-(B-\lambda)y].
\end{eqnarray}

In Eq. (\ref{eq7}), we express $C$, $D$ and $E$ in terms of $A$, $B$
and $F$ by
 Eqs. (\ref{eq8}), (\ref{eq10}) and (\ref{eq11}). After some simplification,
we obtain
\begin{eqnarray}
\setcounter{equation}{12\label{eq12}} aB^2+bB+c=0,
\end{eqnarray}
in which
\begin{align}
a&=1+y^2,\nonumber\\
b&=A-s-2\lambda y^2+2Fxy,\nonumber\\
c&=(\lambda y-Fx)^2+[(A-\lambda)x+Fy]^2+F^2+r-A(s-A).\nonumber
\end{align}
Then $B$ can be solved as
\begin{eqnarray}
\setcounter{equation}{13\label{eq13}}
B=\frac{-b\pm\sqrt{b^2-4ac}}{2a}.
\end{eqnarray}
Hence if we know the values of $A$ and $F$, the value of $B$ is
determined by the Eq. (\ref{eq13}).

\vspace{5mm}
 APPENDIX C
 \vspace{3mm}

  In this appendix, we give the analytical expressions of the
solutions for the quark sector.

  In the quark sector, let
$\lambda=\lambda_3$ and $\alpha=0$. According to Eqs. (\ref{eq6}),
(\ref{eq8}) and (\ref{eq9}), we can express $B$, $D$ and $E$ with
$A$, $C$ and $F$ linearly. The Eqs. (\ref{eq5}) and (\ref{eq7}) are
left invariant. We express $B$, $D$ and $E$ in terms of $A$, $C$ and
$F$ in Eqs. (\ref{eq5}) and (\ref{eq7}). Then in  Eqs. (\ref{eq5})
and (\ref{eq7}), only $A$, $C$ and $F$ are present. Therefore by
Eqs. (\ref{eq5}) and (\ref{eq7}), we can solve $A$ and $C$ in terms
of $F$. The solution is displayed as follows
\begin{eqnarray}
\setcounter{equation}{14\label{eq14}}
A=\frac{-b'-\sqrt{b'^2-4a'c'}}{2a'},
\end{eqnarray}
in which
\begin{align*}
a'&=x^4+x^2y^2+2x^2+y^2+1,\nonumber\\
b'&=-2\lambda_3x^4+2Fyx^3-(\lambda_1+\lambda_2+2\lambda_3y^2+2\lambda_3)x^2+(2Fy^3+2Fy)x-\\
&(1+y^2)(\lambda_1+\lambda_2),\nonumber\\
c'&=\lambda_3^2x^4-2\lambda_3Fyx^3+(F^2+F^2y^2+\lambda_3^2y^2+\lambda_1\lambda_3+\lambda_2\lambda_3)x^2-2\lambda_3F(y^3+y)x+\\
&F^2(y^4+2y^2+1)+(y^2+1)\lambda_1\lambda_2,\nonumber
\end{align*}
The other elements of the mass matrix are given as
\begin{eqnarray}
\setcounter{equation}{13\label{eq13}}
 B&=&\frac{-b-\sqrt{b^2-4ac}}{2a},\\
 \setcounter{equation}{10\label{eq10}}
 C&=&s-A-B,\\
 \setcounter{equation}{8\label{eq8}}
 D&=&-(A-\lambda_3)x-Fy,\\
 \setcounter{equation}{9\label{eq9}}
 E&=&-Fx-(B-\lambda_3)y.
\end{eqnarray}

\vspace{5mm}
  APPENDIX D
\vspace{3mm}

  In this appendix, we give the analytical expressions of the
solutions for the lepton sector.

  In the lepton sector, the analysis in Appendix B and C applies
similarly. Let $\lambda=\lambda_3$ and $\alpha=0$. We find that the
solutions we need can be analytically expressed as follows
\begin{eqnarray}
\setcounter{equation}{15\label{eq15}}
A=\frac{-b'+\sqrt{b'^2-4a'c'}}{2a'},
\end{eqnarray}
in which $b',~a',~c'$ are expressed as the same as that in Appendix
C. Eqs. (\ref{eq8}), (\ref{eq9}), (\ref{eq10}) and (\ref{eq13}) in
Appendix C  apply similarly.
\end{appendix}

\end{document}